\documentclass[intlimits,twoside,a4paper]{article}

\usepackage{amsmath,amssymb}
\usepackage{graphicx}

\usepackage[T2A]{fontenc}
\usepackage[cp1251]{inputenc}

\usepackage[eqsecnum]{cmpj2}
\issue{2014}{17}{1}{13801}
\doinumber{10.5488/CMP.17.13801}

\title[Transition from one- to two-mode generation regime in STNO]%
      {Transition from one- to two-mode generation regime in spin-torque nano-oscillator mediated \\ by thermal noise}
\author[D.V. Slobodianiuk]{D.V. Slobodianiuk}
\address{Taras Shevchenko National University of Kyiv, 64 Volodymyrs'ka St., 01601 Kyiv, Ukraine}

\date{Received June 26, 2013, in final form July 29, 2013}
\authorcopyright{D.V. Slobodianiuk, 2014}

\begin{document}

\maketitle

\begin{abstract}
Two-mode model of spin-torque nano-oscillator (STNO) under the action of thermal noise is considered. Langevin equations for mode amplitudes were derived starting from general nonlinear oscillator model.  Stationary probability distribution function describing mean mode generation powers was obtained using Fokker-Planck equation. It was shown that thermal noise can lead to two-mode generation in STNO. An increase of thermal noise power leads to excitation of the second mode in a system and to a two-mode generation regime through intermediate state when  two modes coexist only in some range of the applied currents.
\pacs 84.40.-x, 84.30.Ng, 85.75.-d
\end{abstract}

\vspace{5mm}
\section{Introduction}

Spin-torque nano-oscillators (STNO) \cite{Kisilev_STO,Rippard} are perfect candidates for creating nanoscale magnetic elements and devices. Interaction of spin polarized current with magnetic moment of such devices leads to destabilization of the magnetic moment orientation \cite{Berger}. So far,  such systems provide the highest precision angle of magnetization.  High amplitude of magnetic precision in STNO leads to a variety of different nonlinear effects in these systems. For example, frequency of magnetization precision gradually depends on the amplitude of precision~--- STNOs are essentially nonisochronic \cite{Tutor}.

Another striking feature of STNOs is their small size. Therefore, the effect of thermal noise on their dynamics is of a critical importance. Moreover, subcritical regime of STNO, when the applied current is smaller than threshold ($I<I_\textrm{th}$), cannot be analyzed without taking thermal noise into account \cite{Noise_1}.
Spectral linewidth of the STNO modes and their dependence on temperature and STNO geometry also require thermal noise effects to be taken into consideration in such systems \cite{Tutor}.

One of the nonlinear effects that can be detected in such devices is a two-mode generation regime when magnetization of the free layer precesses at two frequencies simultaneously. This regime can exist within the range of the applied currents ($I_1<I<I_2$), while below and above this region, a simple one-mode generation regime is established in a system. At the moment, there is no clear understanding of the processes leading to this regime. Practical realization of the regime opens up vast opportunities for STNO application in the applied nanoelectronics and tunable spin-wave generators.

Two possible mechanisms of such a behavior were proposed: mode coupling through non-resonant parametric processes \cite{Melkov_Ch} or thermal noise mediated mode-hopping \cite{Akerman_Ch}. The goal of the current work is to investigate the effect of thermal noise on the dynamics of STNO. The effect of thermal noise on the two-mode generation regime will be investigated using the Fokker-Planck equation formalism for mean-mode powers. The main results of the proposed theory show that an increase of thermal noise level is capable of effectively mediating a two-mode generation regime in STNO.

\section{Theoretical model}

Mode dynamics in such a system can be described using a coupled equation of the so-called nonlinear oscillator model for complex amplitudes of the modes $c_i$ having the form \cite{Kabos,Tutor}:
\begin{eqnarray}
\label{init}
%\begin{array}{l}
&& \frac{{\rd{c_0}}}{{\rd t}} + \ri{{\tilde \omega }_0}({c_0,c_1}){c_0} + {{\tilde \Gamma }_0}({c_0},{c_1}){c_0} = 0,\nonumber\\
&& \frac{{\rd{c_1}}}{{\rd t}} + \ri{{\tilde \omega }_1}({c_0,c_1}){c_1} + {{\tilde \Gamma }_2}({c_0},{c_1}){c_1} = 0.
% \end{array}
\end{eqnarray}
Complex mode amplitude $c_i$ is related to a mode power as $p_i=|c_i|^2$ and a mode precision phase $\phi=\mathrm{arg}(c)$. It should be noted that the modes $c_0$ and $c_1$ are related to the system geometry. Mode $c_0$ corresponds to the lowest frequency eigenmode of the free magnetic layer. For example, in the case of normally magnetized cylindrical layer, this is a uniform mode having indexes ($n,l$)=(0,0) ($n$ is radial mode index and $l$ is azimuthal mode index \cite{Naletov}). Mode $c_1$ is eigenmode of the system having the frequency next to the uniform one. In the mentioned case of normally magnetized cylindrical free layer, this is mode ($n,l$)=(0,1) \cite{Naletov}.

Equations (\ref{init}) have a general form that takes into consideration  different nonlinearities of STNO. Nonlinear frequency shift has the following form:
\begin{eqnarray}
\label{w}
&& \tilde \omega_0=\omega_0+T |c_0|^2+T|c_1|^2, \nonumber\\
&& \tilde \omega_1=\omega_1+T |c_0|^2+T|c_1|^2,
\end{eqnarray}
where $T$ is nonlinear frequency shift coefficient. Typical values of this coefficient can be estimated as $T \simeq {\omega _M}/2$, where $\omega_M=4\pi\gamma M_0$. Here, $\gamma$ is gyromagnetic ratio for electron spin and $M_0$ is saturation magnetization of the sample. For simplicity, in equation (\ref{w}) it is assumed that nonlinear frequency shifts are the same for both modes. Generally speaking, these shifts gradually depend on the mode profile and can be different for different modes \cite{Naletov}. Initial mode frequencies $\omega_0$, $\omega_1$ correspond to the eigenfrequencies of the free magnetic layer as it was discussed before.

The action of spin-polarized current $I$ applied to the system and leading to a decrease of the modes dissipation is accounted for as follows \cite{Tutor}:
\begin{align}\label{dissip}
 {{\tilde \Gamma }_0} &=\Gamma_0^{+}-\Gamma_0^{-}= {\Gamma _0}\left[ {1 - \Delta \left( {1 - \alpha |c_0|^2 - \beta |c_1|^2} \right)} \right]  ,\nonumber\\
 {{\tilde \Gamma }_1} &=\Gamma_1^{+}-\Gamma_1^{-} = {\Gamma _1}\left[ {1 - \frac{{{\Gamma _0}}}{{{\Gamma _1}}}\Delta \left( {1 -\alpha |c_1|^2 - \beta |c_0|^2} \right)} \right], \nonumber\\
 \Delta&=I/I_\textrm{th}\,.
\end{align}
Here, $I_\textrm{th}$ is the value of critical current corresponding to the excitation of the first mode $c_0$. Thus, a dimensionless variable $\Delta$ is normalized in such a way that $\Delta=1$ corresponds to the excitation of the first mode. $\Delta$ has a physical meaning of supercriticality.

Some simple features of the equations (\ref{init}) can be obtained neglecting a nonlinear frequency shift $T=0$. The threshold for mode $c_0$ excitation by current is $\Delta_0=1$
and for the mode $c_1$~--- $\Delta_1=\frac{(\Gamma_1 / \Gamma_0) \alpha-\beta}{\alpha-\beta}$.
In a case $\alpha=\beta$, the mode $c_0$ will be exited at $\Delta_0=1$, while the mode $c_1$ will not be excited at all. Stationary solution for mode powers is \cite{Tutor}:
\begin{eqnarray}\label{sol1}
&&|c_0|^2=p_0=\frac{\Delta-1}{\alpha \Delta}\,,\nonumber\\
&&|c_1|^2=p_1=0\,.
\end{eqnarray}
Regime (\ref{sol1}) can be naturally referred to as one-mode generation regime. Our goal is to investigate how thermal noise in a system affects the one-mode generation regime (\ref{sol1}). Moreover, solution (\ref{sol1}) is invalid in case $\Delta<1$. In fact, this regime corresponds to the regenerative regime of STNO and can be described by properly taking thermal noise effects into account.

Taking thermal noise into account  leads to the following Langevin-like equations:
\begin{eqnarray}\label{init2}
&& \frac{{\rd{c_0}}}{{\rd t}} + \ri{{\tilde \omega }_0}({c_0,c_1}){c_0} + {{\tilde \Gamma }_0}({c_0},{c_1}){c_0} = f_0(t),\nonumber\\
&& \frac{{\rd{c_1}}}{{\rd t}} + \ri{{\tilde \omega }_1}({c_0,c_1}){c_1} + {{\tilde \Gamma }_1}({c_0},{c_1}){c_1} = f_1(t),
\end{eqnarray}
here, $f_i(t)$ is a random white Gaussian process describing thermal noise in a system. Correlation function of this process can be written in the  form \cite{Noise_1}:
\begin{equation}\label{cor}
\langle f_i(t)^*f_i(t{'}) \rangle =2D_i\delta(t-t{'}),
\end{equation}
where $D_i$ is a diffusion coefficient characterizing the noise amplitude. The expression for this coefficient was obtained in work \cite{Noise_1}, where it was shown that in the case of nonlinear oscillator it should be power-dependent and should have the following form:
\begin{equation}\label{diff}
D_i(p_1,p_2)=\Gamma_i^{+}(p_1,p_2) \frac{k_B T}{\lambda \tilde \omega_i(p_1,p_2)}\,,
\end{equation}
where $\lambda$ is a constant which defines the relation between the system energy and frequency:
\begin{equation}\label{lambda}
E(p_1,p_2)=\lambda \int_0^{p_1}\int_0^{p_2} [\tilde\omega_0(q_1,q_2)+\tilde \omega_1(q_1,q_2)]\mathrm{d}q_1\mathrm{d}q_2\,.
\end{equation}
This constant is phenomenological because estimation of its value requires information on the effective volume of the magnetic material of
the free layer involved in auto-oscillation \cite{Noise_1}.

\section{Fokker-Planck equation formalism for coupled-mode model}

To obtain the mean-mode generation powers $p_i$ under the action of thermal noise, it is convenient to introduce the so-called probability distribution function $P(t,p_0,p_1,\phi_1,\phi_2)$ of the system. As one can see, this function depends not only on the mode powers but also on the mode generation phases $\phi_i$. However, usually there is no need to analyze the full-scale probability distribution function of a system. Stationary probability distribution function $P_0(p_0,p_1)$ is introduced instead. It does not depend on oscillatory phases as shown in \cite{Noise_1}.

Our goal is to separate the amplitude and phase noise. Thus, we introduce substitution $c_i=a_i\exp(\ri\phi_i)$ to the equation (\ref{init2})
leading to the following system:
\begin{eqnarray}\label{amp_phase}
&& \frac{{\rd{a_0}}}{{\rd t}} +(\Gamma_0^{+}-\Gamma_0^{-})a_0 = \textrm{Re}\left[f(t)\exp(-\ri \phi)\right], \nonumber\\
&& \frac{{\rd{\phi_0}}}{{\rd t}} + \omega_0 = \textrm{Im}\left[\frac{f(t)}{a_0}\exp(-\ri \phi)\right]
\end{eqnarray}
and adjoint equations for $a_1$, $\phi_1$. This system is equivalent to the following:
\begin{eqnarray}\label{amp_phase2}
&& \frac{{\rd{a_0}}}{{\rd t}} +(\Gamma_0^{+}-\Gamma_0^{-})a_0 - D_1/a_0 = \sqrt{D_1}f_a(t),\nonumber\\
&& \frac{{\rd{\phi_0}}}{{\rd t}} + \omega_0 = -\sqrt{D_1}f_{\phi}(t)/a_0, \nonumber\\
&& \langle f_a(t) f_a(t{'}) \rangle =2\delta(t-t{'}),\nonumber\\
&& \langle f_{\phi}(t) f_{\phi}(t{'}) \rangle =2\delta(t-t{'})
\end{eqnarray}
as these Langevin equations lead to the same drift and diffusion coefficients \cite{Risken} (in the Stratonovitch definition) and, therefore, describe the same physical system. As one can see, the amplitude evolution does not depend on the phase in such representation. This allows us to neglect phase fluctuations in order to determine the mean mode generation powers. By contrast, the analysis of phase noise in such a system requires information on the mode powers because noise evolution depends on amplitude evolution. Moreover, the neglect of a small nonlinear term $D_1/a_0$ leads to an Ornstein-Uhlenbeck process for amplitude fluctuations.

Fokker-Planck equation for equations (\ref{amp_phase2}) has the following form:
\begin{equation}\label{FP1}
 \frac{\rd{P}}{\rd t}=\frac{\rd}{\rd p_0} \left[2p_0\left(\Gamma_0^{+}-\Gamma_0^{-}\right)\right]+\frac{\rd}{\rd p_0}\left(2p_0 D_0 \frac{\rd P}{\rd p_0}\right)+
 \frac{\rd}{\rd p_1} \left[2p_1\left(\Gamma_1^{+}-\Gamma_1^{-}\right)\right]+\frac{\rd}{\rd p_1}\left(2p_1 D_1 \frac{\rd P}{\rd p_1}\right),
\end{equation}
where the dependence on oscillation phases $\phi_i$ is neglected.

A simple solution of equation (\ref{FP1}) can be found in case of potential conditions \cite{Risken} which leads to the condition:
\begin{equation}\label{PC}
\frac{\rd}{\rd p_1}\left(\Gamma_0^{+}-\Gamma_0^{-}\right)=\frac{\rd}{\rd p_0}\left(\Gamma_1^{+}-\Gamma_1^{-}\right).
\end{equation}
In our case $\Gamma_i^{+}$ does not depend on the mode powers and $\Gamma_i^{-}$ is given by (\ref{dissip}) which leads to:
\begin{eqnarray}\label{PC2}
&& \frac{{\rd{\Gamma_0^{-}}}}{{\rd p_1}} =-\Gamma_0\Delta \beta,\nonumber\\
&& \frac{{\rd{\Gamma_1^{-}}}}{{\rd p_0}} =-\Gamma_0\Delta \beta.
\end{eqnarray}
Thus, potential conditions are fulfilled in our case.

Stationary probability distribution function can be found in our case in the following factorized form \cite{Risken}:
\begin{eqnarray}\label{FP}
&&P_0(p_0,p_1)=N \re^{-\Phi_1(p_0,p_1)}\re^{-\Phi_2(p_0,p_1)}, \nonumber \\
&&\Phi_i=\chi^{-1} \int_0^{p_i} \omega_i(q,p_j)\left[1-\frac{\Gamma_i^{-}(q,p_j)}{\Gamma_i^{+}}\right] \,\mathrm{d}q ,
\end{eqnarray}
here, $\chi$ is a constant proportional to the temperature in a system and $N$ is normalization constant given by condition:
\begin{equation}\label{int}
 \int_0^\infty \int_0^\infty P(p_0,p_1) \,\mathrm{d}p_0\mathrm{d}p_1=1.
\end{equation}

Note that without the action of spin-polarized current ($\Gamma_i^{-}=0$) solution (\ref{FP}) reduces to a simple Boltzmann exponential distribution.
Finally, using a stationary probability function it is possible to calculate the mean-mode powers using a standard formalism:
\begin{eqnarray}\label{mean}
&& \bar{p}_0=\int_0^\infty \int_0^\infty p_0 P(p_0,p_1) \,\mathrm{d}p_0\mathrm{d}p_1\,,\nonumber\\
&& \bar{p}_1=\int_0^\infty \int_0^\infty p_1 P(p_0,p_1) \,\mathrm{d}p_0\mathrm{d}p_1\, .
\end{eqnarray}

\begin{figure}[htb]
\centerline{
\includegraphics[width=0.55\textwidth]{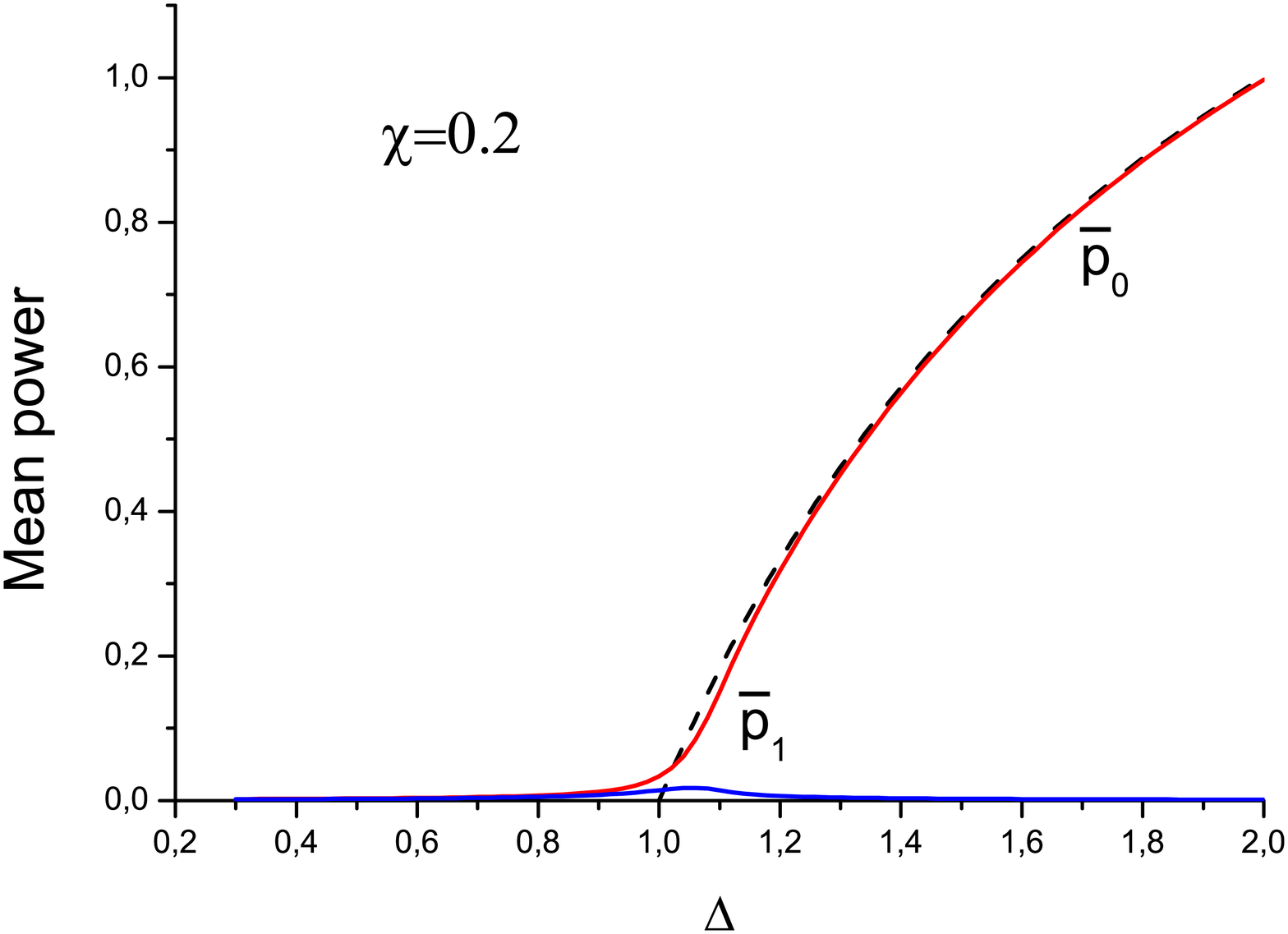}
}
\caption{(Color online) Mean-mode powers dependence on supercriticality $\Delta$. Dashed line corresponds to the one-mode solution (\ref{sol1}).} \label{fig-1}
\end{figure}

\section{Results and discussion}

Using the stationary probability distribution function (\ref{FP}) and equations (\ref{mean}) it is possible to calculate the mean-mode powers under the action of thermal noise in a system. The following parameters were used: $4\pi\gamma M_0=9500$~G, $\Gamma_0=60$~MHz, $\Gamma_1=1.05\Gamma_0$. Integrals in (\ref{mean}) were calculated numerically.

\begin{figure}[htb]
\centerline{
\includegraphics[width=0.55\textwidth]{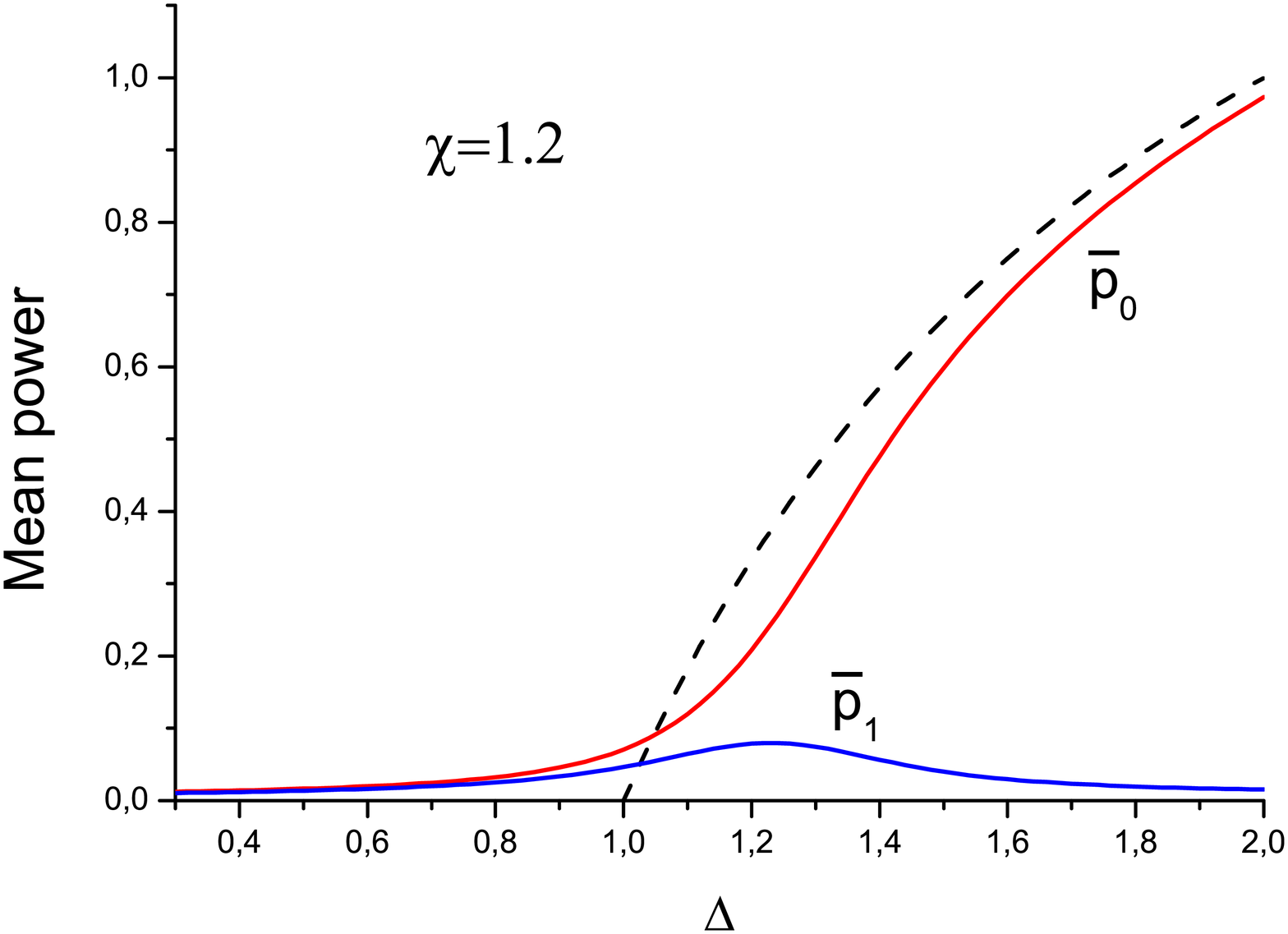}
}
\caption{(Color online) Mean mode powers dependence on supercriticality $\Delta$. Dashed line corresponds to the one-mode solution (\ref{sol1}).} \label{fig-2}
\end{figure}

Figure~\ref{fig-1} shows the dependence of the mean-mode powers for parameter $\chi=0.2$. As one can see in the case of a small power of thermal noise, the mean power of $c_0$ mode practically
does not change compared to the solution without thermal noise (dashed line). The second mode power is insignificantly small in this case.
An increased power of thermal noise leads to various interesting effects shown in figure~\ref{fig-2}. First of all, it should be noted that $c_0$ mode power has nonzero value even in the region $\Delta<1$. As it was discussed before, this case corresponds to the regime when STNO works as regenerative amplifier. The second feature is excitation of mode $c_1$ at some region of the applied currents. This result can be threatened as a two-mode generation regime in a system.

\begin{figure}[htb]
\centerline{
\includegraphics[width=0.55\textwidth]{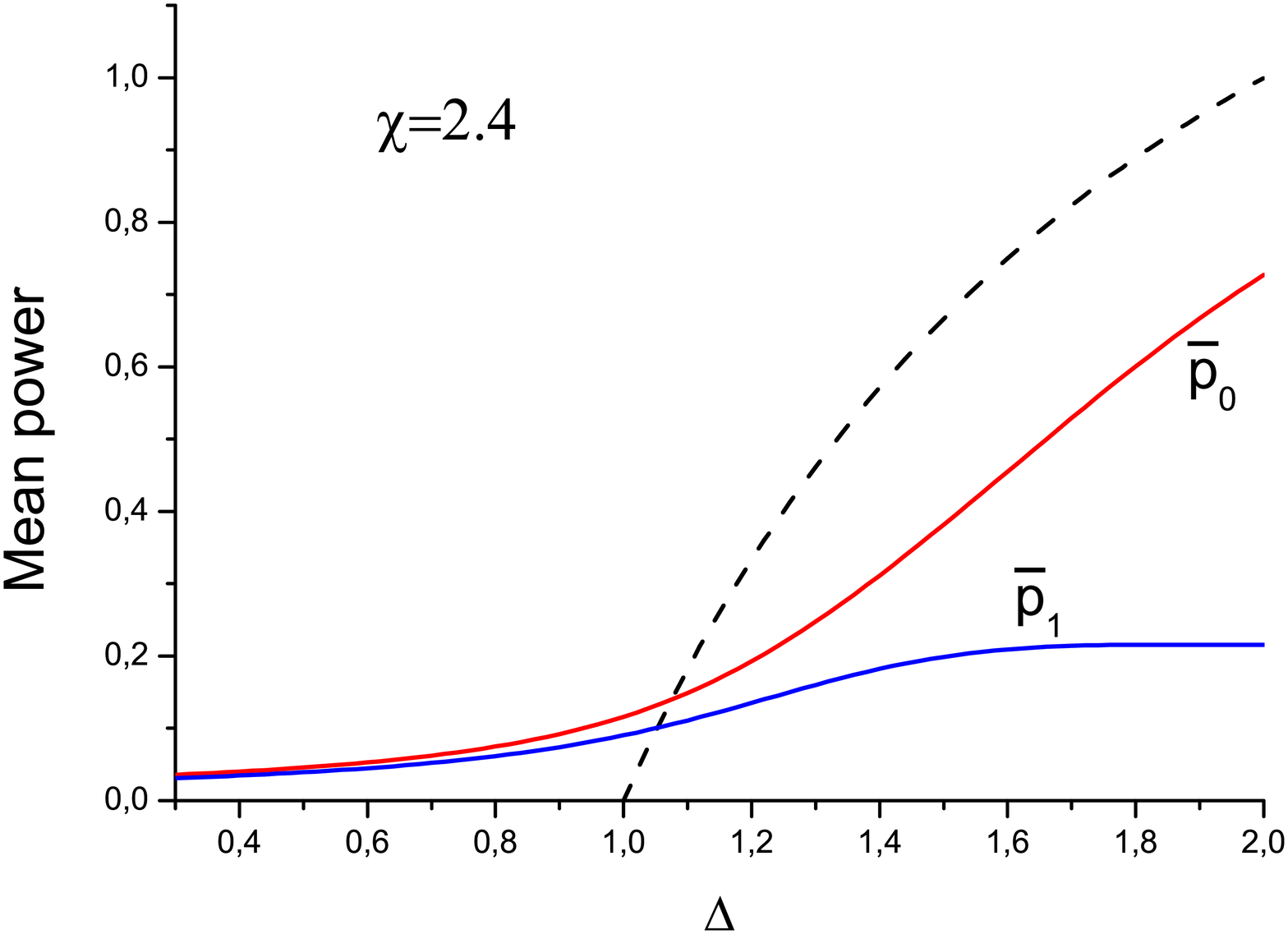}
}
\caption{(Color online) Mean mode powers dependence on supercriticality $\Delta$. Dashed line corresponds to the one-mode solution (\ref{sol1}).} \label{fig-3}
\end{figure}

Finally, figure~\ref{fig-3} shows the results in  case $\chi=2.4$. As one can see, both $c_0$ and $c_1$ modes generate simultaneously in a wide range of the applied current. One should also note how mode $c_0$ power differs from the ideal case (dashed line). This is a result of energy conservation, i.e., system (\ref{init}) can be treated as conservative, having one of integrals $Q=|c_0|^2+|c_1|^2$ which is the total mode power. Obviously, in the case when one has a two-mode generation regime, each mode power will be lower than in the  case of one-mode regime.

It should be noted that figures~\ref{fig-1}--\ref{fig-3} show stationary mode powers. Time evolution of a system is not considered, that is why the present approach is not suitable for analyzing the mode time evolution under the action of spin-polarized current.

\section{Conclusions}

In conclusion, it was demonstrated that thermal noise acting on STNO under the action of spin-polarized current can lead to numerous fundamental effects. When the power of thermal noise is relatively small, one-mode generation regime in STNO is not affected though subcritical regime ($\Delta<1$) appears. This regime corresponds to the regenerative regime when STNO efficiently filters the lowest eigenmode of the STNO from the noise, providing a measurable output microwave signal before the actual threshold of excitation $I=I_\textrm{th}$ for this mode is reached. An increase of thermal noise power leads to the regime at which both modes of the system have nonzero powers at some region of the applied currents. This regime is similar to the one observed in the experiments of two-mode generation regime. The region of two-mode generation is edged by a simple one-mode generation regime where only the lowest frequency mode has nonzero power.

%
%% If you have problems with typesetting in ukrainian uncomment lines below.
%
%  \lastpage
%  \end{document}
\newpage
\ukrainianpart

\title{Перехід від одно- до двомодового режиму генерації в магнітних наноконтактах під дією теплового шуму}
\author{Д.В. Слободянюк}
\address{Київський Національний університет ім. Тараса Шевченка, вул. Володимирська,~64, 01601~Київ, Україна}

%
%% якщо автор є один або автори є з однієї установи:
%
%  \author{1й Автор, 2й Автор, \ldots}
%  \address{Інститут\ldots}
%
%%

\makeukrtitle

\begin{abstract}
\tolerance=3000%
Розглянуто модель магнітного наноконтакту під дією теплового шуму. Виходячи з рівнянь узагальненого нелінійного осцилятора, одержано рівняння Ланжевена для амплітуд мод.
Використовуючи рівняння Фокера-Планка, було одержано вираз для стаціонарної функції розподілу, що описує середні значення потужностей мод системи. Було показано, що тепловий шум в системі може призводити до двомодової генерції. Збільшення потужності теплового шуму призводить до збудження другої моди в системі і встановлення двомодового режиму генерації через проміжний стан, коли в системі існує двомодовий режим генерації в обмеженій області величин зовнішніх струмів.
\keywords магнітні нанонкотакти, спіновий транспорт, тепловий шум
\end{abstract}

\end{document}